\documentclass{aipproc}

\usepackage[dvips]{hyperref}

\hypersetup{%
pdftitle={Physics of EtaPrime->Pi+Pi-Eta and EtaPrime->Pi+Pi-Pi0 decays},
pdfsubject={Hadron Physics},
pdfauthor={Benedykt R. Jany -- Jagiellonian University, Forschungszentrum-J\"ulich},
pdfkeywords={Hadronic Decays of EtaPrime, U(1) axial anomaly, WASA-at-COSY},
bookmarksnumbered,
pdfstartview={FitV},
}%
\layoutstyle{6x9}
\begin{document}

\title{Physics of $\eta'\rightarrow\pi^{+}\pi^{-}\eta$ and $\eta'\rightarrow\pi^{+}\pi^{-}\pi^{0}$ decays}

\classification{13.25.-k}
\keywords      { Hadronic Decays of $\eta'$, $U(1)$ axial anomaly, WASA-at-COSY}

\author{Benedykt R. Jany\thanks{e-mail:~\texttt{b.jany@fz-juelich.de, jany@if.uj.edu.pl}}}{
  address={Nuclear Physics Division of the Jagiellonian University, 30059 Cracow POLAND\newline
 IKP-2 Forschungszentrum-J\"ulich, 52428 J\"ulich GERMANY}
}

\begin{abstract}
The article describes experimental status of the
$\eta'\to\pi^{+}\pi^{-}\eta$ and $\eta'\to\pi^{+}\pi^{-}\pi^{0}$
decays.  A theoretical framework used for description of the decays
mechanism is also reviewed. The possibilities for the measurements with
WASA-at-COSY are mentioned.
\end{abstract}

\maketitle

\section{Experimental overview}
The  $\eta'\rightarrow\pi^{+}\pi^{-}\eta$ decay  with  branching ratio
(BR)  $44.5\%$   \cite{pdg2006}  is  the   main  decay  mode   of  the
$\eta'(958)$ meson. The experimental studies of the reaction mechanism
are  based   on  the  following  data  samples:   $1400$  events  from
\cite{EtaPr1}~(1974),  $6700$ events  from  \cite{EtaPr2} (2000).  The
most  recent results  are from  VES experiment  with two  data sets of
14600 and  7000 events collected  in two different  $\eta'$ production
reactions        \cite{EtaPr3}        (2006).         The        decay
$\eta'\to\pi^{+}\pi^{-}\pi^{0}$  is rare  since  it violates  isospin.
There exists  only a  weak limit $BR\leq5\%$  (90\% CL)  obtained long
time ago \cite{rittenberg}.  These  hadronic decays provide a tool for
studies  of  fundamental  symmetries   of  QCD  as  explained  in  the
following.



\section{Symmetries}
The physical
$\eta'$, $\eta$ and $\pi^{0}$ mesons are not pure isospin states
$\widetilde{\eta'}$, $\widetilde{\eta}$ ($I=0$) and
$\widetilde{\pi}^{0}$ ($I=1$), mainly due to existence of the quark
mass term in QCD Hamiltonian, so the transitions between them are
possible.  Considering  only isospin 
$\pi^{0}-\eta$ mixing, related to light quark mass
difference ($m_{d}-m_{u}$), one can write:
\[
\begin{array}{ccc}
      \pi^{0} & = & \cos(\theta_{\pi\eta})\widetilde{\pi}^{0}-\sin(\theta_{\pi\eta})\widetilde{\eta}\\
			& &\\
      \eta& = & \sin(\theta_{\pi\eta})\widetilde{\pi}^{0}+\cos(\theta_{\pi\eta})\widetilde{\eta}
\end{array}
\]
where  $\theta_{\pi\eta}$ is  the mixing  angle between  $\pi^{0}$ and
$\eta$.  The mixing  angle is expressed in terms  of the quark masses:
\[
\sin(\theta_{\pi\eta})=\frac{\sqrt{3}}{4}\frac{m_{d}-m_{u}}{m_{s}-\widehat{m}},
\]
with $\widehat{m}=(m_{d}+m_{u})/2$.   One proposed way  to extract the
mixing angle is in hadronic decays of $\eta'$
\cite{WilczekEtaPr}.  In the paper it is argued that the ratio
$R=\Gamma(\eta^{'}\to\pi^{+}\pi^{-}\pi^{0})/\Gamma(\eta^{'}\to\pi^{+}\pi^{-}\eta)$
can   be   related  to   $\theta_{\pi\eta}$   in   a very  simple   way:
$R=P\sin^{2}(\theta_{\pi\eta})$, where $P$  is the phase-space factor.
To   extract  $\theta_{\pi\eta}$  there   should  be   two  conditions
fulfilled:
\begin{itemize}
 \item The decay $\eta'\rightarrow\pi^{+}\pi^{-}\pi^{0}$ 
should follow via intermediate state $\pi^{+}\pi^{-}\eta$ and subsequent $\eta-\pi^{0}$ mixing.
\item The amplitudes of both decays must be constant over the phase space.
\end{itemize}

Recent analysis  within  Chiral Unitary Approach
\cite{QuarkBugra} shows  that the decay does not follow entirely via
$\eta-\pi^{0}$ mixing. Moreover
the  decay amplitudes are  not constant since 
resonances in the final state interaction are important.   
Therefore the  mixing angle
$\theta_{\pi\eta}$ and  quark mass  difference cannot be  extracted in
this simple way.  In the  following section a  brief overview of  the Chiral
Effective  Field Theory  is  given and  in  particular Chiral  Unitary
Approach as the physical tool to describe the hadronic decays of $\eta'$
is introduced.

\section{Chiral Effective Field Theory}
Chiral   Symmetry  \cite{Scherer,   KochAspects}  is   connected  with
transformation of left  and right handed quark fields.   It holds for
QCD Hamiltonian  when masses  of the  quarks are put  to zero 
(chiral limit).   It is
spontaneously broken i.e the ground state does not posses the symmetry
of the  Hamiltonian itself.  The  consequence is the existence  of the
massless mode,  so called  Goldstone Bosons --  the octet  of the
pseudoscalar mesons. Nevertheless $\eta'$ in chiral limit remains 
massive as a consequence of  $U(1)$ axial
anomaly \cite{U1anomaly} and it is close related to it.  Chiral 
symmetry is also explicitly broken by the
masses  of  the  quarks what  leads  to  the  non-zero masses  of  the
pseudoscalar  mesons.  with help comes Partially  Conserved Axial
Current  (PCAC) hypothesis, which says  that  the  quark mass  scale  
is small  in
comparison to  the hadron mass  scale.  The symmetry  breaking effects
are small and one can use  a perturbative approach at low energy.  One
can use Effective  Field Theory \cite{LepageRenormalize} and construct
the  Chiral  Effective  Field  Theory  -  \textit{Chiral  Perturbation
Theory} (CHPT).

We  know  that low  energy  QCD should  be  independent  of the  short
distance  physics.  So,  one  derives an  effective Hamiltonian  which
consists of:  a long range part  with ultraviolet cutoff  - to exclude
high momentum states  and a set of contact terms -  to mimic the short
distance  behavior. Each such  contact term,  derived by  the symmetry
constrains, consists  of a local coupling  constant multiplied by the local
operator. Now we can mimic with arbitrary precision the low energy
data sets. But  a problem arises when one  want to generate resonances
-- the  perturbative series just  breaks down  and does  not converge.
One possible  way to  generate resonances is  to use  non perturbative
relativistic coupled channels via  Bethe-Salpeter Equation (BSE) as it
is  done in  the  Chiral Unitary  Approach \cite{HadronicBugra}. The 
$\eta'$ meson is included in this approach as a
dynamical degree of freedom explicitly.

\begin{figure}[ht!bp]
\includegraphics[width=\textwidth, angle=0]{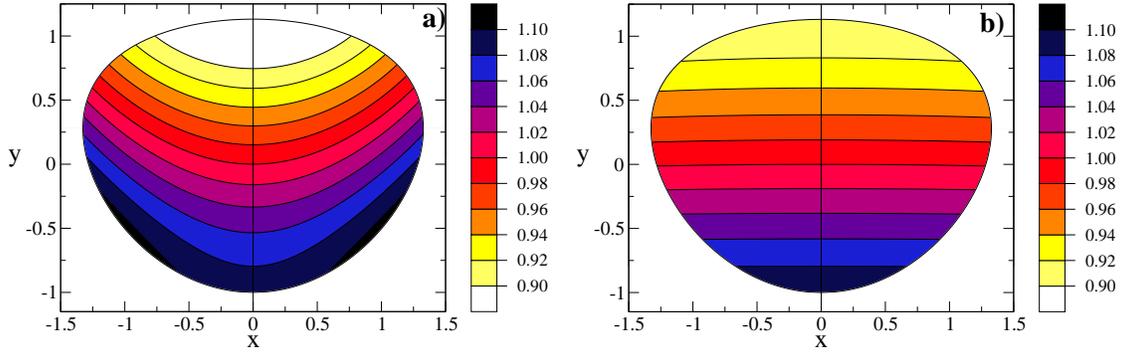}
 \caption{Predicted Dalitz Plot for the $\eta'\rightarrow\pi\pi\eta$ decay: 
a) with $a_{0}(980)$ dominance, b) with $f_{0}(980)/\sigma$ dominance \cite{PrivBugra}.}
\label{Fig:aofo}
\end{figure}

Based on the described framework and on the available data, from which
one has  to extract the  chiral parameters, one can predict  distributions of
the   products    of   the    hadronic   $\eta'$   decays.    In   the
Fig.~\ref{Fig:aofo}  a predicted  shape  of the  Dalitz  plot for  the
$\eta'\to\pi\pi\eta$ decay is shown. Influence of $a_{0}(980)$
($I=1$)    Fig.~\ref{Fig:aofo}a    or   $f_{0}(980)/\sigma$    ($I=0$)
Fig.~\ref{Fig:aofo}b  in the  final  state can  be observed  by different
population of the Dalitz plot.  The limited statistics of the existing
experimental  data  does not  allow  to  distinguish  between the  two
possible scenarios since the  difference is small (the total variation
of the Dalitz plot densities is only 20\%).

\begin{figure}[ht!bp]
\includegraphics[height=0.3\textheight, angle=270]{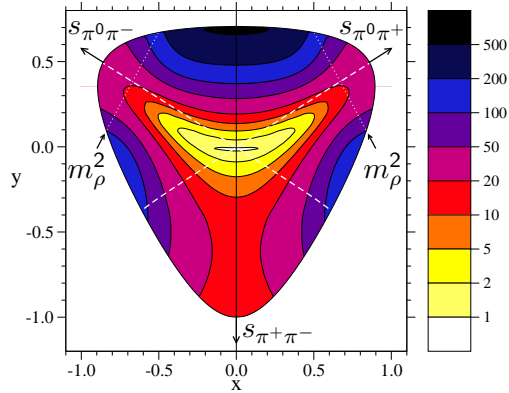}
 \caption{Predicted Dalitz Plot for the $\eta'\rightarrow\pi^{+}\pi^{-}\pi^{0}$,  one sees $\rho^{\pm}(770)$ dominance \cite{HadronicBugra}.}
\label{Fig:rho}
\end{figure}

The fits to  all existing data on hadronic  $\eta$ and $\eta'$ decays,
including the  latest VES data  \cite{EtaPr3}, lead to  predictions of
surprisingly  large branching  ratio  of the  isospin violating  decay
$\eta'\rightarrow\pi^{+}\pi^{-}\pi^{0}$.    The  predicted   value  of
$1.8\%$ is  more than one order  of magnitude larger  than the neutral
mode  $\eta'\rightarrow\pi^{0}\pi^{0}\pi^{0}$.   Such large  branching
ratio is explained by  $\rho^{\pm}(770)$ dominance and can be observed
also  in  the Dalitz  plot  in  Fig.~\ref{Fig:rho}.  The  experimental
search for $\eta'\rightarrow\pi^{+}\pi^{-}\pi^{0}$ is difficult due to
three pion background which accompanies production of $\eta'$.


\section{Summary and Outlook}
The              $\eta'\rightarrow\pi^{+}\pi^{-}\eta$              and
$\eta'\rightarrow\pi^{+}\pi^{-}\pi^{0}$  decays,  as described  above,
give unique  scientific opportunity to  study symmetries in  nature and  
to provide
experimental verification of the sophisticated theoretical predictions
as the Chiral Unitary Approach  and to reveal the driving mechanism of
the decays.   The experimental situation of the  considered decays was
also  reviewed. The  need for  the further  data was  indicated.  When
running    WASA-at-COSY\cite{proposalWASA}     at    its    designed
     luminosity    of
$L=10^{32}\textrm{cm}^{-2}\textrm{s}^{-1}$   one  could   get  $90000$
accepted events  per day for  the $\eta'\rightarrow\pi^{+}\pi^{-}\eta$
decay -- more than the present world statistics.
Newly commissioned WASA-at-COSY experiment is on the way to take exclusive data on $\eta'$ decays.

\bibliographystyle{aipproc}
\bibliography{BRJanyEtaPrime}

\end{document}